\documentclass[%        Class options:
aps,%                   American Physical Society
prd,%                   Physical Review 
tightenlines,%          Single spaced lines
superscriptaddress,%    Authors' addresses linked with superscripts
nofootinbib,%           Does not treat footnotes as references
floatfix]%              Fixes float errors
{revtex4}%              REVTEX 4 Package used
\usepackage{graphicx,%  Default Latex 2eps package for embedding figures
longtable,%              Useful for long table
color%					Color package
}

\begin{document}
\title{Status of three-neutrino oscillation parameters, circa 2013}
\author{		F.~Capozzi}
\affiliation{   	Dipartimento Interateneo di Fisica ``Michelangelo Merlin,'' %\\
               		Via Amendola 173, 70126 Bari, Italy}
\affiliation{   	Istituto Nazionale di Fisica Nucleare, Sezione di Bari, %\\
               		Via Orabona 4, 70126 Bari, Italy}
\author{		G.L.~Fogli}
\affiliation{   	Dipartimento Interateneo di Fisica ``Michelangelo Merlin,'' %\\
               		Via Amendola 173, 70126 Bari, Italy}
\affiliation{   	Istituto Nazionale di Fisica Nucleare, Sezione di Bari, %\\
               		Via Orabona 4, 70126 Bari, Italy}
\author{		E.~Lisi}
\affiliation{   	Istituto Nazionale di Fisica Nucleare, Sezione di Bari, %\\
               		Via Orabona 4, 70126 Bari, Italy}
\author{		A.~Marrone}
\affiliation{   	Dipartimento Interateneo di Fisica ``Michelangelo Merlin,'' %\\
               		Via Amendola 173, 70126 Bari, Italy}
\affiliation{   	Istituto Nazionale di Fisica Nucleare, Sezione di Bari, %\\
               		Via Orabona 4, 70126 Bari, Italy}
\author{		D.~Montanino}
\affiliation{   	Dipartimento  di Matematica e Fisica ``Ennio De Giorgi,'' %\\
               		Via Arnesano, 73100 Lecce, Italy}
\affiliation{   	Istituto Nazionale di Fisica Nucleare, Sezione di Lecce, %\\
               		Via Arnesano, 73100 Lecce, Italy}
\author{		A.~Palazzo}
\affiliation{ 		Max-Planck-Institut f\"ur Physik (Werner Heisenberg Institut), 
                F\"ohringer Ring 6, 80805 M\"unchen, Germany
}
%\date{{\today}}

\begin{abstract}
The standard three-neutrino ($3\nu$) oscillation framework is being increasingly refined by results coming
from different sets of experiments, using neutrinos from solar, atmospheric, 
accelerator and reactor sources. At present, each of the known oscillation parameters 
[the two squared mass gaps  $(\delta m^2,\,\Delta m^2)$ 
and the three mixing angles $(\theta_{12},\,\theta_{13},\,\theta_{23})$] is dominantly 
determined by a single class of experiments. 
Conversely, the unknown parameters [the mass hierarchy, the $\theta_{23}$ octant 
and the CP-violating phase $\delta$] can be currently constrained only 
through a combined analysis of various (eventually all) classes of experiments. In the light of recent new
results coming from reactor and accelerator experiments, and of their interplay with solar and atmospheric data, 
we update the estimated $N\sigma$ ranges of the known $3\nu$ parameters, and revisit the status of the unknown
ones. Concerning the hierarchy, 
no significant difference emerges between normal and inverted mass ordering. 
\textcolor{black}{A slight overall preference is found for  $\theta_{23}$ in the first octant and for nonzero CP violation with $\sin\delta<0$; however, for both parameters, such preference exceeds $1\sigma$ only for normal hierarchy.}
We also discuss the correlations and stability of the oscillation parameters within different combinations of data sets.
\end{abstract}
\pacs{14.60.Pq, 13.15.+g, 11.30.Er} 
\maketitle

%%%%%%%%%%%%%%%%%%%%%%%%%%%%%%%%%%%%%%%%%%%%%%%%%%%%%%%%%%%%%%%%%%%%%%
%%%% Section I %%%%%%%%%%%%%%%%%%%%%%%%%%%%%%%%%%%%%%%%%%%%%%%%%%%%%%%
%%%%%%%%%%%%%%%%%%%%%%%%%%%%%%%%%%%%%%%%%%%%%%%%%%%%%%%%%%%%%%%%%%%%%%

\section{Introduction}

The vast majority of experimental results on neutrino flavor oscillations converge towards   
a simple three-neutrino ($3\nu$) framework, where the flavor states  $\nu_\alpha=(\nu_e,\nu_\mu,\nu_\tau)$ 
mix with the massive states $\nu_i=(\nu_1,\,\nu_3,\,\nu_3)$  via three mixing angles $(\theta_{12},\theta_{13},\theta_{23})$ and
a possible CP-violating phase $\delta$ \cite{Na12}. The observed oscillation frequencies are governed by 
two independent differences between the squared masses $m^2_i$, which 
can be defined as $\delta m^2=m^2_2-m^2_1>0$ and 
$\Delta m^2=m^2_3-(m^2_1+m^2_2)/2$, 
where $\Delta m^2>0$ and $<0$ correspond to normal hierarchy (NH) and inverted hierarchy (IH), respectively \cite{Fo06}.
At present we know five oscillation parameters, each one with an accuracy largely
dominated by a specific class of experiments, namely:
$\theta_{12}$ by solar data, $\theta_{13}$ by short-baseline (SBL) reactor data, $\theta_{23}$ by atmospheric data, mainly from Super-Kamiokande (SK), 
$\delta m^2$ by long-baseline reactor data from KamLAND (KL), and $\Delta m^2$ by long-baseline (LBL) accelerator data, mainly 
from MINOS and T2K.
However, the available data are not yet able to determine the mass hierarchy, to discriminate the $\theta_{23}$ octant, or
to discover CP-violating effects. A worldwide research program is underway to address such open questions
and the related experimental and theoretical issues \cite{Nu12}.

In this context, global neutrino data analyses \cite{Ours,Go12,Go13,Va12} may be useful to get the most restrictive bounds on the known parameters, via the synergic combination of results from different classes of oscillation searches. At the same time, such analyses 
may provide some guidance  about the unknown oscillation parameters, a successful example being represented
by the hints 
of $\sin^2\theta_{13}\sim 0.02 $ \cite{NOVE,HINT,Baha,Ve09}, which
were discussed before the discovery of $\theta_{13}>0$ at reactors \cite{Daya,RENO,DCho}. 
Given the increasing interest on the known oscillation parameters, as well as on possible hints about the unknown ones, 
we find it useful to revisit the previous analysis in \cite{Ours}, 
by including new relevant data which have become available \textcolor{black}{recently (2013--2014)}, and 
which turn out to have an interesting impact on the fit results.

In particular, with respect to \cite{Ours},
we include the recent SBL reactor data from Daya Bay \cite{DY13} and RENO \cite{RE13}, 
which reduce significantly the range of $\theta_{13}$. 
%%%%%%%%%%%%%
\textcolor{black}{We also include the latest 
appearance and disappearance event spectra published in 2013 and at the beginning of 2014
by the LBL accelerator experiments T2K \cite{T2Ka,T2Kd,TK14} and MINOS \cite{MINa,MINd}, which not only constrain
the known parameters 
$(\Delta m^2,\,\theta_{23},\,\theta_{13})$ but, in combination with other data, provide some guidance on the $\theta_{23}$ 
octant and on leptonic CP violation. 
To this regard,  we find 
a slight overall preference for $\theta_{23}<\pi/4$ and for nonzero CP violation with 
$\sin\delta<0$; however,  for both parameters, such hints exceed $1\sigma$ only for normal hierarchy.
No significant preference emerges for normal versus inverted hierarchy.} 
%%%%%%%%%%%%%%
Among the various fit results which can be of interest, we find it useful
to report both the preferred $N\sigma$ ranges of each oscillation parameter and the covariance plots of
selected couples of parameters, as well as to discuss their stability
and the role of different data sets in the global analysis. 

Our work is structured as follows: In Sec.~II we discuss some methodological issues concerning the analysis
of different data sets and their combination. In Sec.~III and IV we
present, respectively, the updated ranges on single oscillation parameters, and the
covariances between selected couples of parameters. We pay particular attention to the 
(in)stability and (in)significance of various hints about unknown parameters, 
also in comparison with other recent (partial or global)
data analyses. Finally, we summarize our work in Sec.~V.

\section{Methodology}
\vspace{-2mm}

In this Section we briefly discuss the various data sets used and how they are combined in the global fit. 

\vspace*{-3.5mm}
\subsection{LBL Acc.\ + Solar + KL data}
\vspace*{-1.5mm}

Concerning LBL accelerator data, we include the observed
energy spectra of events, in both appearance (muon-to-electron flavor) and 
disappearance (muon-to-muon flavor) oscillation modes, as presented by the  
T2K \cite{T2Ka,T2Kd,TK14} and MINOS \cite{MINa,MINd,MI13,MI14} experiments. The theoretical spectra are calculated through
a suitably modified version of the GLoBES software package \cite{GLOB,GLO2}. 
We have verified that our fits reproduce very well the regions allowed at various C.L.\ in \cite{T2Ka,T2Kd,MINa,MINd,MI13,MI14}, 
under the same restrictive assumptions made therein on specific oscillation parameters (e.g., by limiting their range
or fixing them a priori). However, we emphasize that no restrictions
are applied in the global fit discussed in the next Section, where all the $3\nu$ parameters are
free to float. 

At the current level of accuracy, LBL accelerator data (disappearance plus appearance)
are known to be sensitive not only to the dominant parameters $(\pm\Delta m^2,\, \theta_{23}, \,\theta_{13})$, 
but also to the subdominant parameters ($\delta m^2,\,\theta_{12}$) and $\delta$. For this reason, as argued in 
\cite{Ours}, it is convenient to analyze LBL accelerator data in combination with solar and KL data, 
which provide the necessary input for ($\delta m^2,\,\theta_{12}$). We remark that ``Solar + KL'' data 
(here treated as in \cite{Ours}) provide a preference for $\sin^2\theta_{13}\sim 0.02$ in our analysis, 
which plays a role in the combination ``LBL Acc.\ + Solar + KL,'' as discussed in the next Section.

\vspace*{-3.5mm}
\subsection{Adding SBL reactor data}
\vspace*{-1.5mm}

After the recent T2K observation of electron flavor appearance, the combination  of LBL Acc.~+ Solar~+~KL data
can provide a highly significant measurement of $\theta_{13}$ which, however, is somewhat correlated with two unknowns affecting 
LBL data: the CP violating phase $\delta$ and the $\theta_{23}$ octant. It is thus 
important to add the accurate and $(\delta,\,\theta_{23}$)-independent 
measurement of $\theta_{13}$ coming from SBL reactor experiments, within
a  ``LBL Acc.\ + Solar + KL + SBL Reac.'' combination.
In this work, SBL reactor neutrino
data are statistically treated as in \cite{Pa13}, with the further inclusion of the most recent data from Daya Bay \cite{DY13}
and RENO \cite{RE13}.

\vspace*{-3.5mm}
\subsection{Adding atmospheric neutrino data}
\vspace*{-1.5mm}

In this work, the analysis of SK atmospheric neutrino data (phases I--IV) \cite{PhDT,We12,Hi13} is 
essentially unchanged with respect to \cite{Ours}. We remind the
reader that such data involve a very rich oscillation phenomenology which is sensitive, in principle,
also to subleading effects related to the mass hierarchy, the $\theta_{23}$ octant and the CP phase $\delta$ \cite{Malt}. 
However, within the current experimental and theoretical uncertainties, it remains difficult to
disentangle and probe such small effects at a level exceeding $\sim 1\sigma$--$2\sigma$ \cite{Fo06}. Moreover, 
independent $3\nu$ fits of SK~I-IV data \cite{Ours,Go13,Hi13} converge on some but not all the hints about
subleading effects, as discussed later. 
Therefore, as also argued in \cite{Ours}, we prefer to add these data only in the final  ``LBL Acc.\ + Solar + KL + SBL Reac.\ + SK Atm.'' 
combination, in order to separately gauge their effects on the various $3\nu$ parameters.

\vspace*{-3.5mm}
\subsection{Conventions for allowed regions}
\vspace*{-1.5mm}

In each of the above combined data analyses, the six oscillation parameters 
$(\Delta m^2,\, \delta m^2,\, \theta_{12}, \,\theta_{13}, \,\theta_{23})$ are left free
at fixed hierarchy (either normal or inverted). Parameter ranges at $N$ standard deviations
are defined through $N\sigma=\sqrt{\chi^2-\chi^2_{\min}}$. As in \cite{Ours}, this definition is maintained also 
in plots involving two parameters, where it is understood that the previous $N\sigma$ ranges
are reproduced by projecting the two-dimensional contours over one parameter axis \cite{Na12}. It is
also understood that, in each figure, all undisplayed parameters are marginalized away.

Finally, we shall also report the relative preference of the data for either NH or IH, as measured by
the quantity $\Delta \chi^2_{\mathrm{{I}-{N}}}=\chi^2_{\min}(\mathrm{IH})-\chi^2_{\min}(\mathrm{NH})$.
This quantity cannot immediately be translated into ``$N\sigma$'' by taking the square root of its absolute value, 
because it refers to two discrete hypotheses, not connected by variations of a physical parameter.
We shall not enter into the current debate about the statistical interpretation of 
$\Delta \chi^2_{\mathrm{{I}-{N}}}$ \cite{Ba13,Ca13,Bl13} 
because, as shown in the next Section, its numerical values are not yet significant
enough to warrant a dedicated discussion.

\vspace*{-3.5mm}
\section{Ranges of oscillation parameters}
\vspace*{-2mm}

In this Section we graphically report the results of our global analysis of increasingly richer data sets, grouped in accordance
to the previous discussion. 

Figures~1, 2 and 3 show the $N\sigma$ curves for the data sets defined in Sec.~II~A, II~B and II~C, respectively.
In each figure, the solid (dashed) curves refer to NH (IH); however, only the NH curve is shown 
for the $\delta m^2$ and $\theta_{12}$ parameters, since the very tiny effects related to
the NH-IH difference \cite{Ours,Quas} are unobservable in the fit. [Also note that 
the $\delta m^2$ and $\theta_{12}$ constraints change very little in Figs.~1--3.]  
For each parameter in Figs.~1--3, the more linear and symmetrical are the curves, 
the more gaussian is the probability distribution 
associated to that parameter.  

Figure~1 refers to the combination LBL Acc.\ + Solar + KL, which already sets 
(without the need of
atmospheric and reactor data) highly significant lower and upper bounds on all the oscillation parameters, 
except for $\delta$. In this figure, the relatively strong appearance
signal in T2K \cite{T2Ka} plays an important role: it dominates the lower bound on $\theta_{13}$, and also drives
the slight but intriguing preference for $\delta\simeq 1.5\pi$, since for $\sin\delta \sim -1$ the CP-odd term in the
$\nu_\mu\to\nu_e$ appearance probability \cite{Cerv,Freu} is maximized \cite{T2Ka}. This trend wins over
the current MINOS preference for $\sin\delta \gtrsim 0$ \cite{MINa,MI14}, since the T2K appearance signal is stronger 
than the  MINOS one and dominates in the global fit. 
%%%%%%%%
\textcolor{black}{On the other hand, MINOS disappearance data \cite{MINd,MI14} still 
lead to a slight preference for nonmaximal $\theta_{23}$, as compared with nearly maximal $\theta_{23}$ in the T2K data fit
\cite{T2Kd,TK14}. The (even slighter) preference for the second $\theta_{23}$ octant is due to the interplay of LBL accelerator and
Solar + KL data, as discussed in the next Section.}
%%%%%%%%%%%%%%

Figure~2 shows the results obtained by adding (with respect to Fig.~1) the SBL reactor data, 
whose primary effect is a strong reduction of the $\theta_{13}$ uncertainty. Secondary effects
include: ($i$) a slightly more pronounced preference for $\delta \simeq 1.5\pi$ and $\sin\delta<0$,
and ($ii$) a swap of the preferred $\theta_{23}$ octant with the hierarchy ($\theta_{23}<\pi/4$ in NH and $\theta_{23}>\pi/4$ in IH). 
These features will be interpreted in terms of parameter covariances in the next Section. 

Figure~3 shows the results obtained by adding (with respect to Fig.~2) the SK atmospheric data in the most complete data set.
It thus represents a synopsis of the current constraints on each oscillation parameter, according to our global $3\nu$ analysis. 
The main  differences with respect to Fig.~2 include: $(i)$ an even more pronounced preference for $\sin\delta <0$, with
a slightly lower best fit at $\delta\simeq 1.4\pi$; $(ii)$  a slight reduction of the  errors on $\Delta m^2$ and a relatively
larger variation of its best-fit value with the hierarchy; $(iii)$ a preference for $\theta_{23}$ in the first octant
for both NH and IH, which is a persisting feature of our analyses \cite{Fo06,Ours}.  
The effects $(ii)$ and $(iii)$  show that atmospheric neutrino data have the potential to probe subleading
hierarchy effects, although they do not yet emerge in a stable or significant way. 
Concerning the effects $(i)$, it should be noted that the
existing full $3\nu$ analyses of atmospheric data \cite{Go13,Hi13}, as well as this work, consistently show
that such data prefer $\delta$ around $1.5\pi$ or slightly below, although with still large uncertainties.  
Table~I summarizes in numerical form the results shown in Fig.~3.

When comparing Figs.~1--3, it is interesting to note an increasingly pronounced preference for nonzero CP violation
with increasingly rich data sets, although the two CP-conserving  cases $(\delta =0,\,\pi)$ remain 
allowed at $\lesssim 2\sigma$ in both NH and IH, even when all data are combined (see Fig.~3). 
It is worth noticing that the two maximally CP-violating cases ($\sin\delta = \pm1$) have opposite likelihood:
while the range around $\delta\sim 1.5\pi$ ($\sin\delta\sim-1$) is consistently preferred,  small
ranges around $\delta \sim 0.5\pi$ ($\sin\delta\sim +1$) appear to be disfavored (at $>2\sigma$ in Fig.~3). 
In particular, for the specific case of NH and at $\sim 90\%$ C.L.\ ($\sim 1.6\sigma$), only
the range $\sin\delta <0$ is allowed in Fig.~3, while the complementary one is disfavored, with the two CP-conserving
cases being just ``borderline.'' 
In the next few years, the appearance channel in LBL accelerator experiments will provide crucial 
data to investigate these intriguing CP violation hints.

From the comparison of Figs.~1--3 one can also notice a slight overall preference for nonmaximal
mixing $(\theta_{23}\neq 0)$, although it appears to be weaker than in \cite{Ours}, essentially because the most
recent T2K data prefer nearly maximal mixing \cite{T2Kd,TK14}, and thus ``dilute'' the opposite
preference coming from MINOS \cite{MINd,MI14} and atmospheric data \cite{Ours}. Moreover, the indications about the octant
appear to be somewhat unstable in different combinations of data.  In the present analysis, only
atmospheric data consistently prefer the first octant in both hierarchies, but the 
\textcolor{black}{global fit} significance is
non-negligible $(\sim 90\%$~C.L.) only in NH (see Fig.~3). 
\textcolor{black}{By excluding LBL accelerator data from the global fit,
the significance of $\theta_{23}<\pi/4$ would raise to $\sim 2\sigma$ in NH and $\sim 1.5\sigma$ in IH (not shown). 
It should be noted that, in a recent $3\nu$ global fit \cite{Go13}}, 
the preferred octant toggles with the hierarchy, while in the latest atmospheric $3\nu$ 
\textcolor{black}{analyses from the SK collaboration 
\cite{We12,Hi13} (without LBL accelerator data)}
the second octant is preferred in both NH and IH. We remark that such differences in the $\theta_{23}$ fit results should not be
considered as conflicting with each other, 
since they are all compatible within the (still large) quoted uncertainties.

\textcolor{black}{We also emphasize that no atmospheric $\nu$ analysis performed outside the SK collaboration \cite{Ours,Go12,Go13,Va12} 
can possibly reproduce in detail the official SK one, which currently includes hundreds of bins
and $>150$ systematic error sources \cite{PhDT}; on the other hand, this level of complexity 
also hinders the interpretation of subleading effects at the $\sim 1\sigma$ level, such as those related to (non)maximal
mixing, which are diluted over many data points and whose size is comparable to systematic uncertainties. 
We continue to argue, as discussed in \cite{Fo06}, that our slight preference for $\theta_{23}<\pi/4$ in atmospheric $\nu$ data stems
from a small but persisting overall excess of low-energy electron-like events; see also \cite{Go12} for a similar discussion. We are
unable to trace the source of a slight preference for $\theta_{23}>\pi/4$ in the official SK analysis.}  
In any case, these fluctuations in atmospheric fit results show how difficult it is to reduce the allowed range 
of $\theta_{23}$ \textcolor{black}{on the basis of atmospheric neutrino data only}. In this context, 
the disappearance channel in LBL accelerator experiments will provide independent and \textcolor{black}{increasingly accurate} 
data to address the issue of nonmaximal $\theta_{23}$ in the next few years.
 
%%%%%%%%%%%
\textcolor{black}{
Finally, we comment on the size of 
$\Delta \chi^2_{\mathrm{{I}-{N}}}$ which, by construction, is not apparent in Figs.~1--3. 
We find $\Delta \chi^2_{\mathrm{{I}-{N}}}=-1.4,\,-1.1,\,-0.3$, for the data sets in Figs.~1, 2, and 3, 
respectively. Such values are both small and decreasing with increasingly rich data sets; thus, they 
do not provide us with relevant indications about the  hierarchy.}
%%%%%%%%%%%%%% 

\newpage
\phantom{.}
\vspace*{1cm}
\phantom{.}
\begin{figure}[h]
\includegraphics[width=0.95\textwidth]{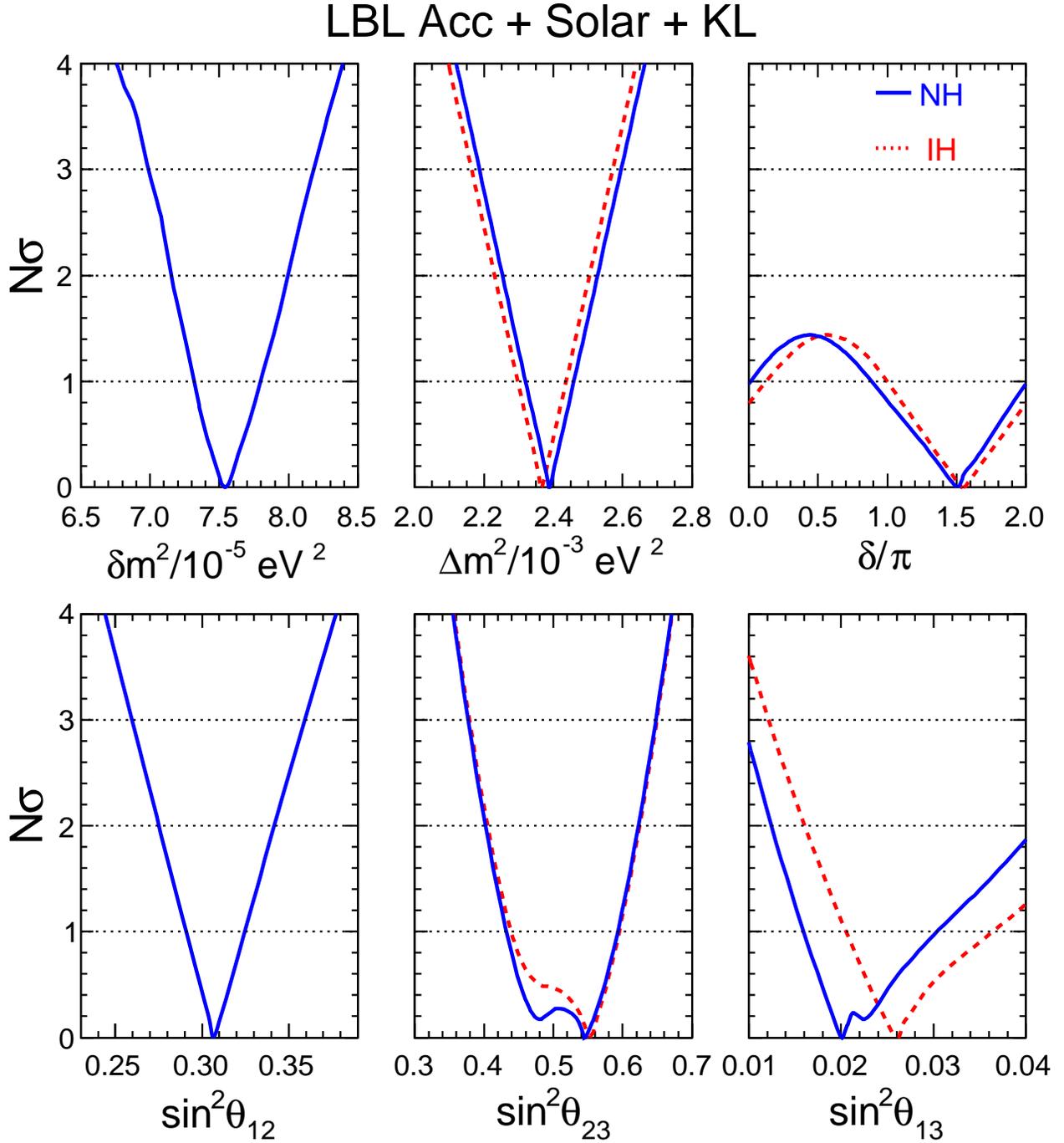}
\caption{\label{fig1} Combined $3\nu$ analysis of LBL Acc.\ + Solar + KL data: Bounds on the oscillation parameters in terms
of standard deviations $N\sigma$ from the best fit. Solid (dashed) lines refer to NH (IH). The horizontal dotted lines
mark the $1\sigma$, $2\sigma$ and $3\sigma$ levels for each parameter (all the others being marginalized away). See the text for details.}
\end{figure}
\vfill
\phantom{.}

\newpage
\phantom{.}
\vspace*{1cm}
\phantom{.}
\begin{figure}[h]
\includegraphics[width=0.95\textwidth]{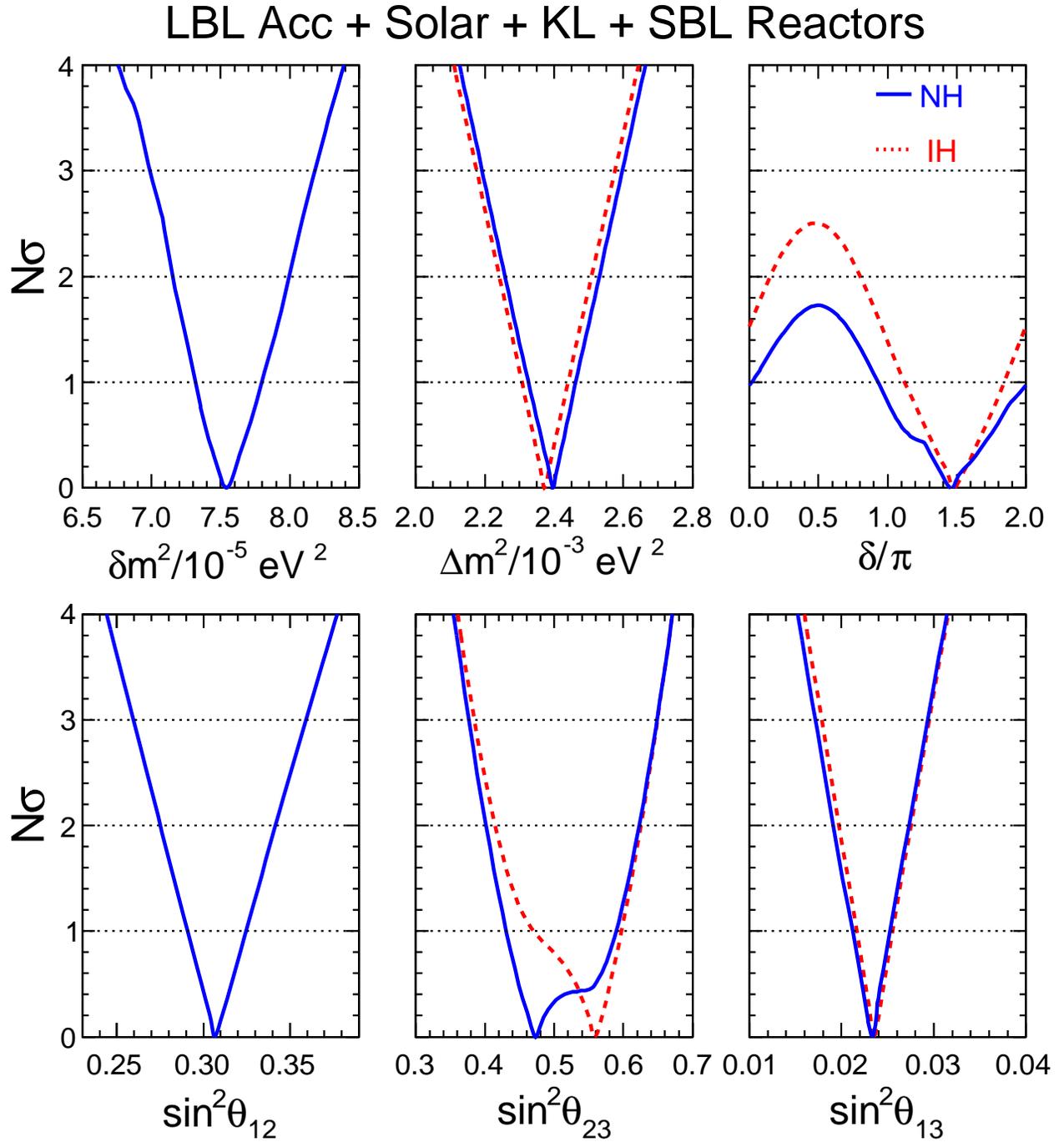}
\caption{\label{fig2} As in Fig.~1, but adding SBL reactor data.}
\end{figure}
\vfill
\phantom{.}

\newpage
\phantom{.}
\vspace*{1cm}
\phantom{.}
\begin{figure}[h]
\includegraphics[width=0.95\textwidth]{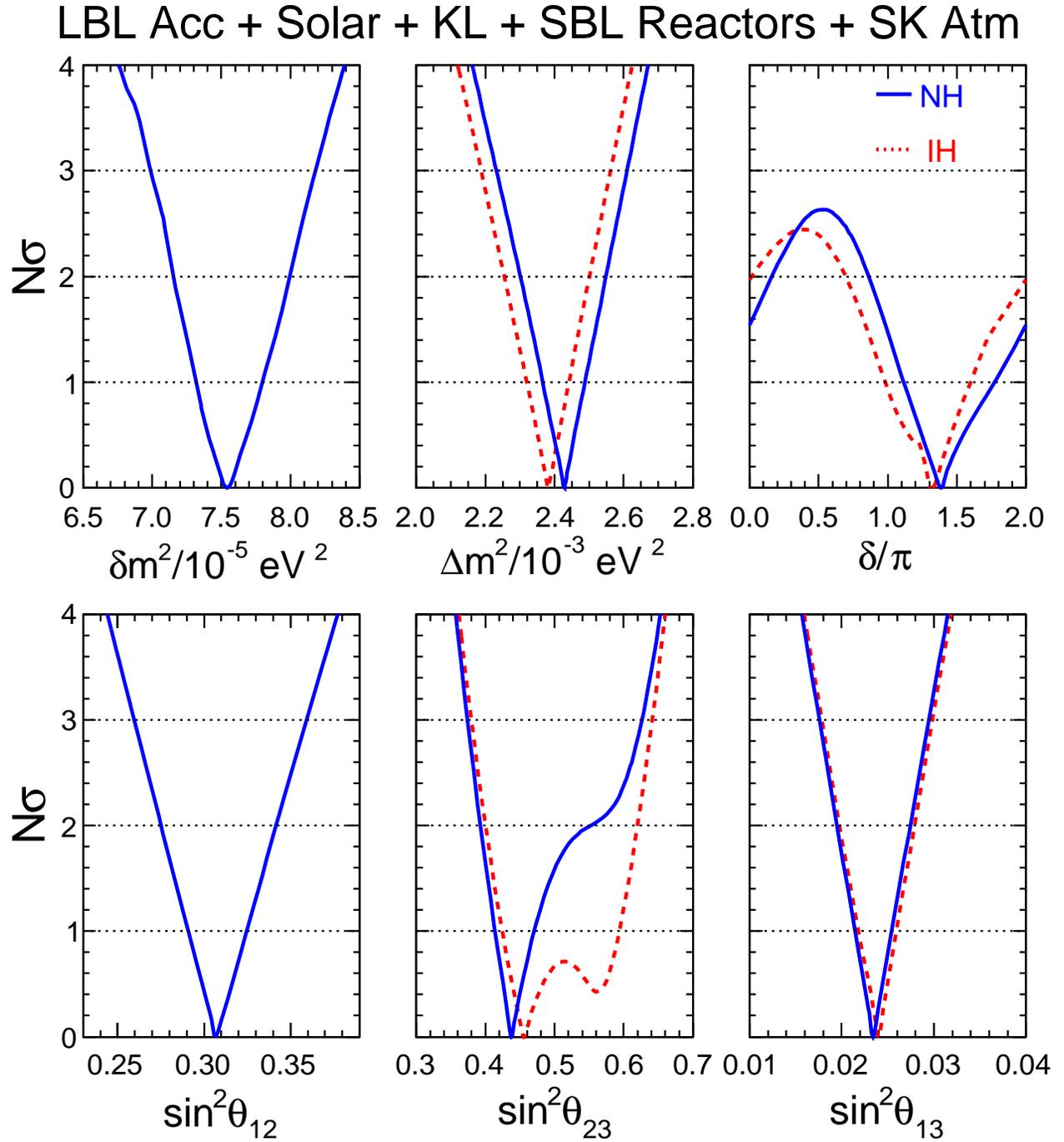}
\caption{\label{fig3} As in Fig.~2, but adding SK atmospheric data  in a global $3\nu$ analysis of all data.}
\end{figure}
\vfill
\phantom{.}

%===========================================================================
\begin{table}[t]
\textcolor{black}{
\caption{\label{Synopsis} Results of the global $3\nu$ oscillation analysis, in terms of best-fit values and
allowed 1, 2 and $3\sigma$ ranges  for the $3\nu$ mass-mixing parameters. See also Fig.~3 for a graphical representation 
of the results. We remind that $\Delta m^2$ is defined
herein as $m^2_3-{(m^2_1+m^2_2})/2$, with $+\Delta m^2$ for NH and $-\Delta m^2$ for IH. The CP violating phase is 
taken in the (cyclic) interval $\delta/\pi\in [0,\,2]$. The overall $\chi^2$ difference between IH and NH is insignificant
($\Delta \chi^2_{\mathrm{{I}-{N}}}=-0.3$).}
%\centering
%\resizebox{\textwidth}{!}{
\begin{ruledtabular}
\begin{tabular}{lcccc}
Parameter & Best fit & $1\sigma$ range & $2\sigma$ range & $3\sigma$ range \\
\hline%---------------------------------------------------------------------
$\delta m^2/10^{-5}~\mathrm{eV}^2 $ (NH or IH) & 7.54 & 7.32 -- 7.80 & 7.15 -- 8.00 & 6.99 -- 8.18 \\
\hline%---------------------------------------------------------------------
$\sin^2 \theta_{12}/10^{-1}$ (NH or IH) & 3.08 & 2.91 -- 3.25 & 2.75 -- 3.42 & 2.59 -- 3.59 \\
\hline%---------------------------------------------------------------------
$\Delta m^2/10^{-3}~\mathrm{eV}^2 $ (NH) & 2.43 & 2.37 -- 2.49 & 2.30 -- 2.55 & 2.23 -- 2.61 \\
$\Delta m^2/10^{-3}~\mathrm{eV}^2 $ (IH) & 2.38 & 2.32 -- 2.44 & 2.25 -- 2.50 & 2.19 -- 2.56 \\
\hline%---------------------------------------------------------------------
$\sin^2 \theta_{13}/10^{-2}$ (NH) & 2.34 & 2.15 -- 2.54 & 1.95 -- 2.74 & 1.76 -- 2.95 \\
$\sin^2 \theta_{13}/10^{-2}$ (IH) & 2.40 & 2.18 -- 2.59 & 1.98 -- 2.79 & 1.78 -- 2.98 \\
\hline%---------------------------------------------------------------------
$\sin^2 \theta_{23}/10^{-1}$ (NH) & 4.37 & 4.14 -- 4.70 & 3.93 -- 5.52 & 3.74 -- 6.26 \\
$\sin^2 \theta_{23}/10^{-1}$ (IH) & 4.55 & 4.24 -- 5.94 & 4.00 -- 6.20 & 3.80 -- 6.41 \\
\hline%---------------------------------------------------------------------
$\delta/\pi$ (NH) & 1.39 & 1.12 -- 1.77 & 0.00 -- 0.16 $\oplus$ 0.86 -- 2.00 &  ---  \\
$\delta/\pi$ (IH) & 1.31 & 0.98 -- 1.60 & 0.00 -- 0.02 $\oplus$ 0.70 -- 2.00 &  ---  \\
\end{tabular}
\end{ruledtabular}
%}%end of resizebox
%\vspace*{.2cm}
}
\end{table}
%============================================================================

\section{Covariances of oscillation parameters}

In this Section we show the allowed regions for selected couples of oscillation parameters, and discuss
some interesting correlations. 

Figure~4 shows the global fit results in the plane charted by 
($\sin^2\theta_{23},\,\Delta m^2$), in terms of regions allowed at 1, 2 and $3\sigma$ ($\Delta\chi^2=1$, 4 and 9).
Best fits are marked by dots, and it is understood that
all the other parameters are marginalized away. From left to right, the panels 
refer to increasingly rich datasets, as previously discussed:
LBL accelerator + solar + KamLAND data (left), plus SBL reactor data (middle), plus SK atmospheric data (right).
The upper (lower) panels refer to normal (inverted) hierarchy.
This figure shows the instability of the $\theta_{23}$ octant discussed above, in a graphical format which
is perhaps more familiar to most readers. It is worth noticing the increasing ($\sin^2\theta_{23},\,\Delta m^2$)
covariance for increasingly nonmaximal $\theta_{23}$ (both in first and in the second octant), which 
contributes to the overall $\Delta m^2$ uncertainty. In this context, the measurement of $\Delta m^2$ at SBL reactor experiments
(although not yet competitive with accelerator and atmospheric experiments \cite{DY13}) may become relevant in the
future:  being  $\theta_{23}$-independent, it will help to break the current correlation with $\theta_{23}$
and to improve the overall $\Delta m^2$ accuracy in the global fit.

Figure~5 shows the allowed regions in  the plane charted by ($\sin^2\theta_{23},\,\sin^2\theta_{13}$). Let us consider first the 
left panels, where a slight negative correlation between these two parameters emerges from LBL appearance data, as discussed in \cite{Ours}.
The contours extend towards relatively large values of $\theta_{13}$, especially in IH, in order to accommodate
the relatively strong T2K appearance signal \cite{T2Ka}. 
However, solar + KL data  provide independent (although 
weaker) constraints on $\theta_{13}$ and, in particular, prefer $\sin^2\theta_{13}\sim 0.02$ in our analysis. 
\textcolor{black}{This value, being on the ``low side'' of the allowed regions of $\theta_{13}$, leads 
(via anticorrelation) to a best-fit value of $\theta_{23}$ on the ``high side'' (i.e., in the second-octant) for both NH and IH. 
However, when current
SBL reactor data are included in the middle panels, a slightly higher value of $\theta_{13}$ 
is preferred ($\sin^2\theta_{13}\simeq 0.023$) with very small uncertainties: this value is high enough to
flip the $\theta_{23}$ best fit from the second to the first octant in NH, but not in IH.}
 
\textcolor{black}{It is useful to compare the left and middle panels of Fig.~5 with the analogous ones of Fig.~1 from our previous analysis \cite{Ours}: the 
local minima in the two $\theta_{23}$ octants are now closer and more degenerate. This fact is mainly due to
the persisting preference of T2K disappearance data for nearly maximal mixing \cite{TK14}, which is gradually
diluting the MINOS preference for nonmaximal mixing \cite{MI14}. Moreover, accelerator data are becoming increasingly 
competitive with atmospheric data in constraining $\theta_{23}$ \cite{TK14}. Therefore, although we still find (as in previous 
works \cite{Fo06,Ours}) that atmospheric data alone prefer $\theta_{23}<\pi/4$, the overall combination with current non-atmospheric data 
(right panels of Fig.~5) makes this indication less significant than in previous fits (compare, e.g., with Fig.~1 in \cite{Ours}), especially in IH where non-atmospheric data now prefer the opposite case $\theta_{23}>\pi/4$.
The fragility of the $\theta_{23}$ octant fit (with and without atmospheric neutrinos) was also noted in the
recent analysis \cite{Go13}.  
In conclusion, the overall indication for $\theta_{23}<\pi/4$ in both NH and IH (right panels of Fig.~5) 
is currently weaker than in our previous analysis \cite{Ours}; in particular, its significance reaches only
$\sim 1.6\sigma$ ($~90\%$~C.L.) in NH, while it is $<1\sigma$ in IH.
Further accelerator neutrino data will become increasingly important in assessing the status of $\theta_{23}$
in the near future.}

\newpage
%%%%%%%%%%%%%%%%%%%%%%%%%%%%%%%%%%%%%%%%%%%%%%%%%%%%%%%%%%%%%%%%%%%%%
\begin{figure}[t]
\includegraphics[width=0.72\textwidth]{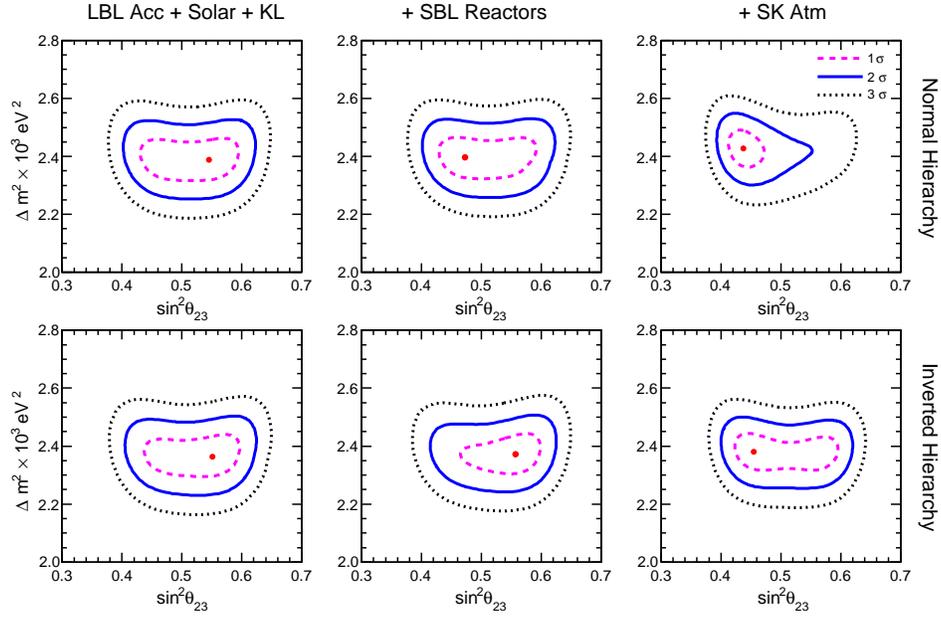}
\caption{\label{fig4} Results of the analysis in the plane charted by 
($\sin^2\theta_{23},\,\Delta m^2$), all other parameters being marginalized
away. From left to right, the 
regions allowed at 1, 2 and $3\sigma$ refer to increasingly rich datasets:
LBL accelerator + solar + KamLAND data (left panels), plus SBL reactor data (middle panels), plus SK atmospheric data (right panels).
Best fits are marked by dots.
The three upper (lower) panels refer to normal (inverted) hierarchy.}
\end{figure}
%%%%%%%%%%%%%%%%%%%%%%%%%%%%%%%%%%%%%%%%%%%%%%%%%%%%%%%%%%%%%%%%%%%%%%%
\phantom{.}
%%%%%%%%%%%%%%%%%%%%%%%%%%%%%%%%%%%%%%%%%%%%%%%%%%%%%%%%%%%%%%%%%%%%%
\begin{figure}[h]
\includegraphics[width=0.72\textwidth]{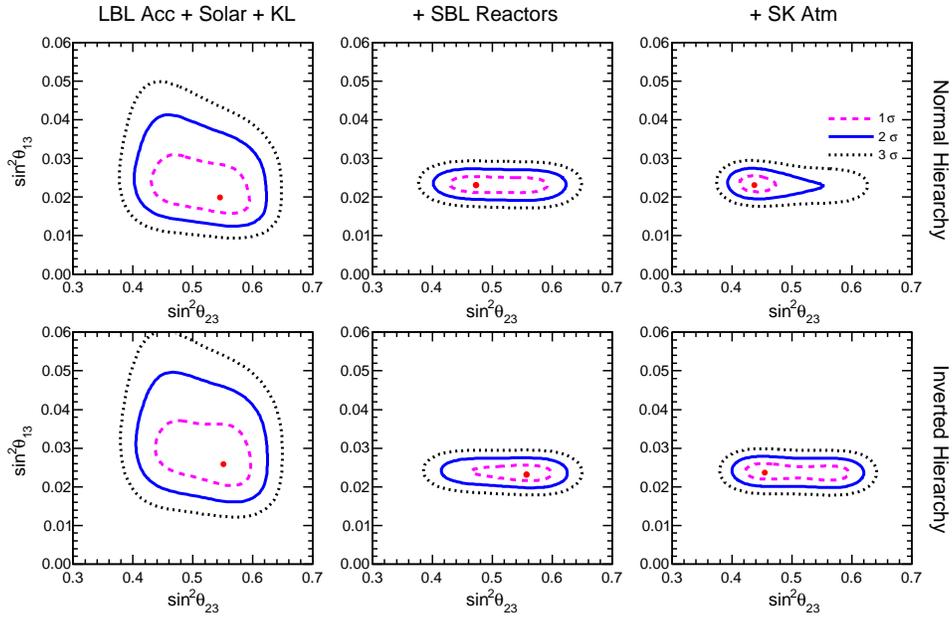}
\caption{\label{fig5} As in Fig.~4, but in the plane  ($\sin^2\theta_{23},\,\sin^2\theta_{13}$).}
\end{figure}
%%%%%%%%%%%%%%%%%%%%%%%%%%%%%%%%%%%%%%%%%%%%%%%%%%%%%%%%%%%%%%%%%%%%%%%

\newpage
%%%%%%%%%%%%%%%%%%%%%%%%%%%%%%%%%%%%%%%%%%%%%%%%%%%%%%%%%%%%%%%%%%%%%
\begin{figure}[t]
\includegraphics[width=0.72\textwidth]{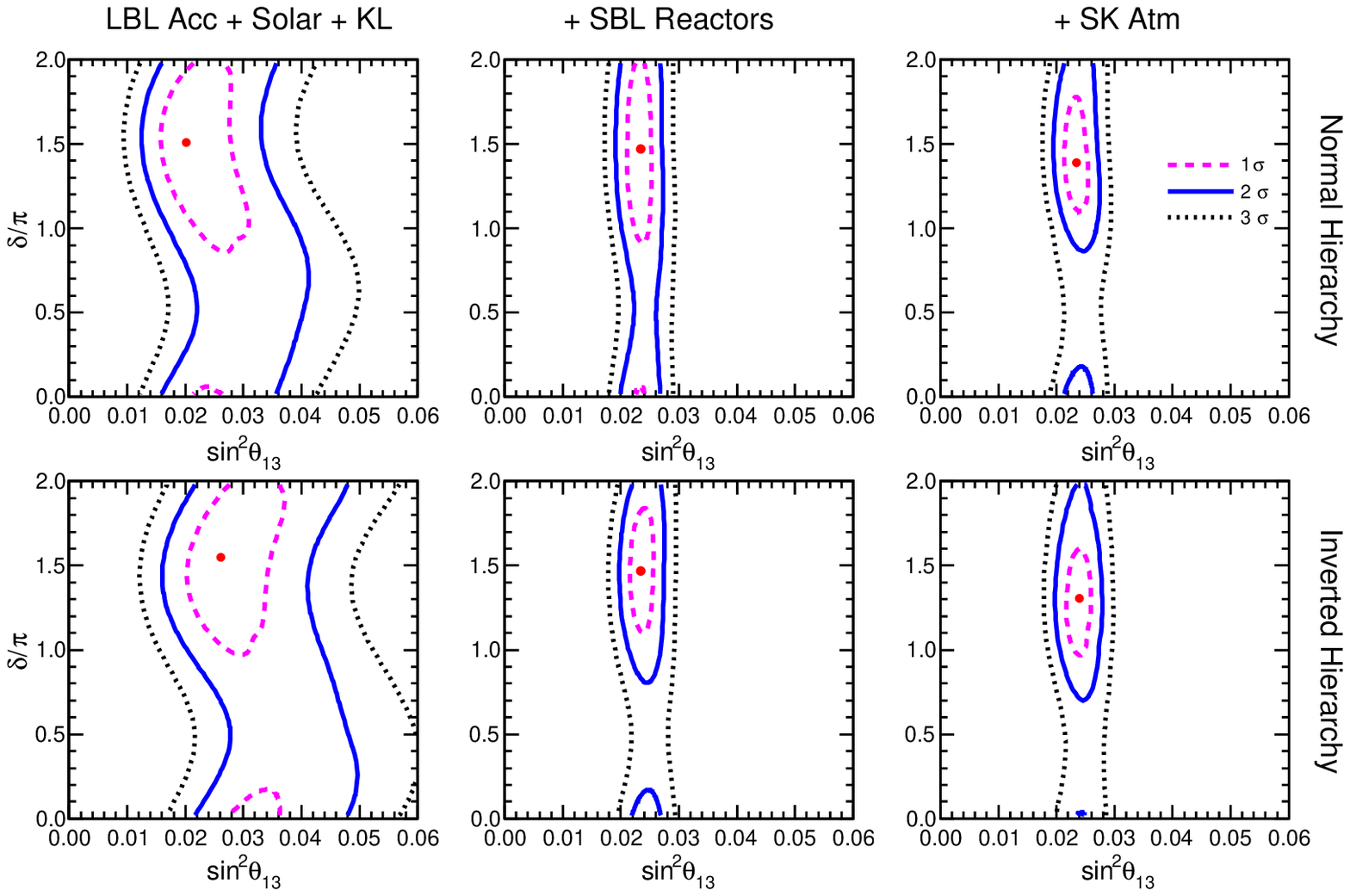}
\caption{\label{fig6} As in Fig.~4, but in the plane  ($\sin^2\theta_{13},\,\delta/\pi$).}
\end{figure}
%%%%%%%%%%%%%%%%%%%%%%%%%%%%%%%%%%%%%%%%%%%%%%%%%%%%%%%%%%%%%%%%%%%%%%%
\phantom{.}
%%%%%%%%%%%%%%%%%%%%%%%%%%%%%%%%%%%%%%%%%%%%%%%%%%%%%%%%%%%%%%%%%%%%%
\begin{figure}[h]
\includegraphics[width=0.72\textwidth]{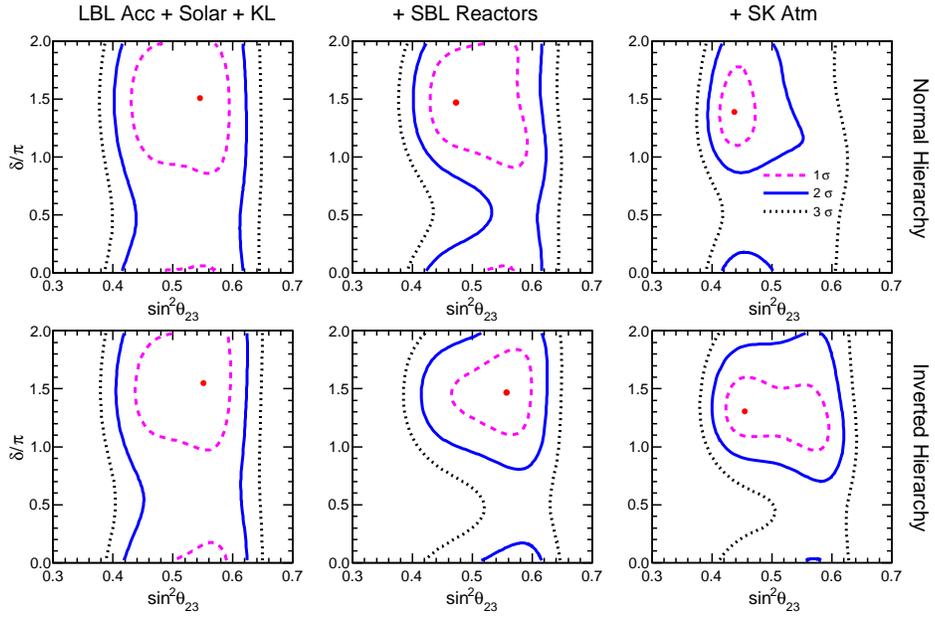}
\caption{\label{fig7} As in Fig.~4, but in the plane  ($\sin^2\theta_{23},\,\delta/\pi$).}
\end{figure}
%%%%%%%%%%%%%%%%%%%%%%%%%%%%%%%%%%%%%%%%%%%%%%%%%%%%%%%%%%%%%%%%%%%%%%%

\newpage
Figure~6 shows  the allowed regions in  the plane ($\sin^2\theta_{13},\,\delta/\pi$), which is
at the focus of current research in neutrino physics. In the left panels, with respect to previous
results in the same plane \cite{Ours}, there is now a more marked preference for $\delta\sim 1.5\pi$, where
a compromise is reached between the relatively high $\theta_{13}$ values preferred by the
T2K appearance signal, and the relatively low value preferred by solar + KL data. In the middle panel,  
SBL reactor data strengthen this trend by reducing the covariance between
$\theta_{13}$ and $\delta$.  It is quite clear that we can still learn much from the combination
of accelerator and reactor data in the next few years. Finally,
the inclusion of SK atmospheric data in the right panels also adds
some statistical significance to this trend, with a slight lowering of the best-fit value 
of $\delta$. 

Figure~7 completes our discussion by showing  the allowed regions in  the plane ($\sin^2\theta_{23},\,\delta/\pi$).
The shapes of the allowed regions are rather asymmetrical in the two $\theta_{23}$ octants, which are physically
inequivalent in the flavor appearance phenomenology of accelerator and atmospheric neutrinos. Therefore,
reducing the octant degeneracy will also help, indirectly, our knowledge of $\delta$. Eventually, more 
subtle covariances may be studied in this plane \cite{Park}, but we are still far from the required accuracy.

\section{Summary and Conclusions}

In the light of recent new
\textcolor{black}{data (circa 2013-2014)} coming from reactor and accelerator experiments, 
and of their interplay with solar and atmospheric data, 
we have updated the estimated $N\sigma$ ranges of the known $3\nu$ parameters ($\Delta m^2,\, \delta m^2,\,
 \theta_{12},\,\theta_{13},\,\theta_{23}$), and we have revisited the status of the current unknowns
[$\mathrm{sign}(\Delta m^2),\,\mathrm{sign}(\theta_{23}-\pi/4),\,\delta$]. The results of the
global analysis of all data are shown in Fig.~3 and in Table~I, from which one can derive the 
ranges of the known parameters; in particular, as compared with a previous analysis \cite{Ours}, 
one can appreciate 
a significant reduction of the $\theta_{13}$ uncertainties, and some changes in the $(\Delta m^2,\,\theta_{23})$
ranges. 

\textcolor{black}{We have also discussed in detail the status of the unknown parameters.
Concerning the hierarchy [$\mathrm{sign}(\Delta m^2)$], we still 
find no appreciable difference between normal and inverted mass ordering. 
With respect to \cite{Ours}, we continue to find an overall preference 
for the first $\theta_{23}$ octant, but with a lower statistical significance, which  
exceeds $1\sigma$ only in NH. This feature of the current analysis is mainly due to the persisting preference of
(increasingly accurate) T2K disappearance data for nearly maximal mixing \cite{TK14},
as opposed to somewhat different indications coming from the analysis of 
MINOS \cite{MI14} and atmospheric data \cite{Ours}. Probably the most intriguing 
feature of the current data analysis is the emergence of an overall preference for nonzero  
CP violation around $\delta \sim 1.4\pi$ (with $\sin\delta <1$) at $\gtrsim 1\sigma$ level,
while some ranges with $\sin\delta>1$ are disfavored at $\gtrsim 2\sigma$.}

In order to understand how the various constraints and hints emerge from the analysis, and to appreciate
their (in)stability, we have considered increasingly rich data sets, starting from the combination of LBL
accelerator plus solar plus KamLAND data, then adding SBL reactor data, and finally including atmospheric data.
We have discussed the fit results both on single parameters and on selected
couples of correlated parameters. \textcolor{black}{We remark that the $\theta_{23}$ octant issue appear somewhat 
unstable at present, while the hints about $\delta$ (despite being still statistically weak) seem to arise from an
overall convergence of several pieces of data.} Of course, these might just be fluctuations: the
search for
[$\mathrm{sign}(\Delta m^2),\,\mathrm{sign}(\theta_{23}-\pi/4),\,\delta$] is still open to all possible
outcomes. In this context, joint $3\nu$ analyses of LBL
accelerator data (in both appearance and disappearance mode) and SBL reactor data have the potential
to bring interesting new results in the next few years.

\acknowledgments
F.C., G.L.F., E.L., A.M., and D.M.\ acknowledge  
support from the Istituto Nazionale di Fisica Nucleare (INFN, Italy) through the ``Astroparticle Physics'' 
research project. 
A.P. acknowledges support from the European Community through a Marie Curie Intra-European 
Fellowship, grant agreement PIEF-GA-2011-299582 ``On the Trails of New Neutrino Properties."
He also acknowledges partial support from the  European Union FP7 ITN INVISIBLES 
(Marie Curie Actions, PITN-GA-2011-289442).\textcolor{black}{
Preliminary results of this work were presented by E.L.\ at {\em NNN13}, XIV International Workshop on Next generation Nucleon Decay and Neutrino Detectors (Kashiwa, Japan, 2013), at {\em NuPhys 2013}, Topical research meeting on Prospects in Neutrino Physics (London,
UK, 2013); and at {\em ICFA 2014}, European meeting of the International Committee for Future Accelerators (Paris, France, 2014).}

\newpage

%%%%%%%%%%%%%%%%%%%%%%%%%%%%%%%%%%%%%%%%%%%%%%%%%%%%%%%%%%%%%%%%%%%%%%%%%%%%%%%%

\end{document}